\documentclass{ifacconf}

\usepackage{graphicx}
\usepackage{natbib}
\usepackage{graphics}
\usepackage{times} 
\usepackage{amssymb}
\usepackage{algorithm}
\usepackage{mathtools}
\usepackage{bm}
\usepackage{xcolor}
\usepackage{arydshln}
\usepackage{microtype}
\usepackage{float}
\usepackage{tikz}
\usepackage{pgfplots}
\pgfplotsset{compat=newest}

\newlength\axisheight
\newlength\axiswidth

\newcommand{\mr}[1]{\mathrm{#1}}

\RequirePackage[absolute]{textpos}
\DeclareRobustCommand*{\copyrightnote}[1]{%
  \begin{textblock}{200}(15,286.2)
      \footnotesize #1%
  \end{textblock}%
}

\begin{document}
\begin{frontmatter}

\title{Distributed Model Predictive Control for Periodic Cooperation of Multi-Agent Systems} 

\thanks[footnoteinfo]{%
F. Allg{\"o}wer and M. A. M{\"u}ller are thankful that this work was funded by the Deutsche Forschungsgemeinschaft (DFG, German Research Foundation) -- AL 316/11-2 - 244600449.
F. Allg{\"o}wer is thankful that this work was funded by the Deutsche Forschungsgemeinschaft (DFG, German Research Foundation) under Germany’s Excellence Strategy -- EXC 2075 -- 390740016.
}

\author[First]{Matthias K{\"o}hler} 
\author[Second]{Matthias A. M{\"u}ller} 
\author[First]{Frank Allg{\"o}wer}

\address[First]{University of Stuttgart, Institute for Systems Theory and Automatic Control, Stuttgart, Germany (e-mails: \{koehler, allgower\}@ist.uni-stuttgart.de).}
\address[Second]{Leibniz University Hannover, Institute of Automatic Control, Hanover, Germany (e-mail: mueller@irt.uni-hannover.de)}

\begin{abstract}
We consider multi-agent systems with heterogeneous, nonlinear agents subject to individual constraints that want to achieve a periodic, dynamic cooperative control goal which can be characterised by a set and a suitable cost.
We propose a sequential distributed model predictive control (MPC) scheme in which agents sequentially
solve an individual optimisation problem to track an artificial periodic output trajectory.
The optimisation problems are coupled through these artificial periodic output trajectories, which are communicated and penalised using the cost that characterises the cooperative goal.
The agents communicate only their artificial trajectories and only once per time step.
We show that under suitable assumptions, the agents can incrementally move their artificial output trajectories towards the cooperative goal,
and, hence, their closed-loop output trajectories asymptotically achieve it. 
We illustrate the scheme with a simulation example.
\end{abstract}

\begin{keyword}
Predictive control, distributed MPC,
multi-agent systems, cooperative control, nonlinear systems
\end{keyword}

\end{frontmatter}
%===============================================================================
\copyrightnote{\copyright 2023 the authors. This work has been accepted to IFAC for publication under a Creative Commons Licence CC-BY-NC-ND.}
\section{Introduction}
Control of a multi-agent system, i.e. a system comprising subsystems, called agents, that can be operated independently of each other, often requires a distributed controller to ensure, e.g. independent operation and scalability with respect to the number of agents.
Common control goals of multi-agent systems include, for example, consensus or synchronisation, formation control, or other objectives which go beyond stabilisation of some \emph{a priori} defined equilibrium, see, e.g.~\cite[]{OlfatiSaber2007,Cao2013,Oh2015}.
Many applications, e.g. cooperating robot systems or multi-vehicle cooperation, consider uncoupled systems connected by a shared objective.
A possible control strategy for these systems is distributed model predictive control (MPC) with its valuable ability to handle explicitly nonlinear systems and constraints.
Various distributed MPC schemes have been designed to control multi-agent systems, see, e.g. \cite[]{Muller.2012,Maestre2014a,Mueller2017b,Nikou2019}.

In this work, we make use of a tracking formulation with artificial references, pioneered for tracking of constant references with linear systems in~\cite[]{Limon.2008}, and further developed for periodic references in~\cite[]{Limon.2016}.
The idea is to track a given external reference by introducing an artificial reference as an additional decision variable.
It can be shown that by suitably penalising the distance of the artificial reference to the external reference, eventually the closed loop converges to a reachable reference closest to the external one.
Subsequently, extensions to nonlinear systems for tracking of constant references~\cite[]{Limon.2018} and of periodic references~\cite[]{Koehler2020b} were proposed.
For a distributed setup,~\cite[]{Ferramosca2011} present a distributed Gauss-Jacobi type scheme for (dynamically coupled) linear systems and~\cite[]{Carron2020a} solve a coverage problem for nonlinear multi-agent systems. 
In~\cite[]{Carron2023}, a distributed MPC scheme with artificial references for multi-agent systems is proposed where connectivity is ensured by imposing a lower bound constraint on the Fiedler eigenvalue of the graph Laplacian.
A practical comparison of a graph theoretic approach and a distributed MPC scheme with artificial references is given in~\cite[]{Ebel2021} for the task of formation control, highlighting the potential of the latter.

In comparison to these works, we do not impose on the multi-agent system an external reference that achieves the cooperative goal, but instead design a sequential distributed MPC scheme that lets the agents coordinate the eventual cooperative trajectory themselves. 
To this end, we use similar ideas as developed in~\cite[]{Koehler2022b}, which proposes a sequential distributed MPC scheme using artificial references to steer a multi-agent system to a self-organised cooperative equilibrium.
In contrast, in this work, we consider the goal of dynamic, periodic cooperation, which requires a modified optimisation problem.
In addition, we do not rely on a specific setting as in~\cite[]{Koehler2022b}, but state more general assumptions to establish asymptotic achievement of the cooperative goal.
The main idea is the following: Each agent is equipped with a local MPC optimisation problem that tracks an artificial output reference, which is also a decision variable.
Next, we introduce a cooperation cost which couples the local optimisation problems through the artificial output references, penalising the distance of the artificial output references to the cooperative goal.
The agents first move their artificial output references towards the cooperative goal, and then asymptotically achieve the cooperative goal themselves.
Due to the use of artificial references, we expect a larger region of attraction than directly tracking trajectories that solve the cooperative problem.
In addition, the proposed scheme can handle agents joining or leaving the system, and requires communication only once per agent in each time step.

\subsection{Notation}
The maximal eigenvalue of a matrix $A = A^\top$ is denoted by $\bar{\lambda}_A$.
The interior of a set $\mathcal{A}$ is denoted by $\mr{int}\,\mathcal{A}$.
The non-negative reals are denoted by $\mathbb{R}_{\ge 0}$, and $\mathbb{N}_0$ denotes the natural numbers including 0.
The set of integers from $a$ to $b$, $a \le b$, is denoted by $\mathbb{I}_{a:b}$.
Given a positive (semi-)definite matrix $A=A^\top$, the corresponding (semi-)norm is written as $\Vert x \Vert_A = \sqrt{x^\top A x}$.
Given a collection of $m$ vectors $v_i \in \mathbb{R}^{n_i}$, $i \in \mathbb{I}_{1:m}$, we denote the stacked vector by $v = \mathrm{col}_{i=1}^m(v_i) = [v_1^\top \dots v_m^\top]^\top$.
%===============================================================================
\section{Multi-agent system}\label{sec:setup}
We consider a multi-agent system comprising $m \in \mathbb{N}$ heterogeneous agents with nonlinear discrete-time dynamics
\begin{subequations}\label{eq:agent_equations}
    \begin{align}
        x_i(t + 1) &= f_i(x_i(t), u_i(t))\label{eq:agent_dynamics}, \\
        y_i(t) &= h_i(x_i(t), u_i(t))\label{eq:output_relation}
    \end{align}
\end{subequations}
with state $x_i(t) \in \mathbb{X}_i \subseteq \mathbb{R}^{n_i}$, input $u_i(t) \in \mathbb{U}_i \subseteq \mathbb{R}^{q_i}$, and output $y_i(t) \in \mathbb{Y}_i \subset \mathbb{R}^{p}$ at time $t\in\mathbb{N}_0$, continuous $f_i: \mathbb{X}_i \times \mathbb{U}_i \to \mathbb{X}_i$ and $h_i: \mathbb{X}_i \times \mathbb{U}_i \to \mathbb{Y}_i$.
We assume that the agents are subject to individual pointwise-in-time constraints $(x_i(t), u_i(t)) \in \mathcal{Z}_i \subset \mathbb{X}_i \times \mathbb{U}_i$ for $t \in \mathbb{N}_0$, where $\mathcal{Z}_i$ is compact.

The agents can communicate according to an undirected graph $\mathcal{G} = (\mathcal{V}, \mathcal{E})$ with vertices $\mathcal{V}$ and edges $\mathcal{E}$.
Each agent is assigned a vertex $i\in\mathcal{V}$ which are connected through edges $e_{ij} = e_{ji} \in\mathcal{E}$.
The set of neighbours of agent $i$ is then $\mathcal{N}_i = \{j \in\mathcal{V} \mid e_{ji} \in \mathcal{E}\}$, i.e. it contains all agents with which agent $i$ may communicate.
In the following, we assume lossless and immediate communication.

The control goal is that the agents' outputs converge to a periodic trajectory which satisfies a cooperative objective, e.g. synchronisation or flocking.
However, the eventual periodic output trajectory is not given \emph{a priori} by an external governor, but could be any that achieves the cooperative goal.
We characterise this cooperative goal through an \emph{output cooperation set} $\mathcal{Y}_T^{\mathrm{c}}$, which contains possible output trajectories that achieve this goal, as further specified in Definition~\ref{def:cooperation_set_and_cost} below. 

%===============================================================================
\section{Distributed MPC scheme}
We begin by defining admissible periodic output trajectories and corresponding trajectories in the states and inputs of the agents.
We assume in the following a fixed period length $T \in \mathbb{N}$.
\begin{assum}\label{asm:reference}
    There exist a non-empty compact set 
    \begin{align*}
        \mathcal{R}_{T,i} \subseteq \{ &r_{T,i} = (x_{T,i}, u_{T,i}) \in \mathcal{R}_{i} \subseteq \mathrm{int}\,\mathcal{Z}_{i}^{T+1} \mid
        \\
        &x_{T,i}(k+1) = f_i(x_{T,i}(k), u_{T,i}(k)), k\in\mathbb{I}_{0:T-1},
        \\
        &x_{T,i}(T) = x_{T,i}(0), u_{T,i}(T) = u_{T,i}(0) \}
    \end{align*}
    and a non-empty, compact and convex set 
    \begin{align}\label{eq:cooperation_output_trajectories_set}
        \mathcal{Y}_{T, i} \subseteq \{ &y_{T,i} \in \mathbb{Y}_i^T \mid \exists r_{T,i} = (x_{T,i}, u_{T,i}) \in \mathcal{R}_{T,i},
        \\
        &y_{T,i}(k) = h_i(x_{T,i}(k), u_{T,i}(k)), k\in\mathbb{I}_{0:T-1} \} \notag
        .
    \end{align}
\end{assum}
A periodic cooperation output trajectory $y_{T,i} \in \mathcal{Y}_{T, i}$ and the corresponding cooperation trajectory $r_{T,i} \in \mathcal{R}_{T,i}$ from~\eqref{eq:cooperation_output_trajectories_set} are called \emph{admissible}.
We write $y_{T,i}(k \vert t)$ and $r_{T,i}(k \vert t)$ to denote the $k$-th step of a cooperation (output) trajectory that is created at time step $t$.
Moreover, we denote by $y_{T,i}(\cdot{+}1 \vert t)$ the trajectory generated from shifting $y_{T,i}(\cdot \vert t)$ by one step, i.e. $y_{T,i}(\cdot{+}1 \vert t) = (y_{T,i}(1 \vert t), y_{T,i}(2 \vert t), \dots, y_{T,i}(T-1 \vert t), y_{T,i}(0 \vert t))$.
With a minor abuse of notation, if $k \ge T$, then $y_{T,i}(k \vert t) = y_{T,i}(k \bmod T \vert t)$ and $r_{T,i}(k \vert t) = y_{T,i}(k \bmod T \vert t)$.
For two periodic cooperation (output) trajectories $y_{T,i}$ and $\hat{y}_{T,i}$, define $d(y_{T,i}, \hat{y}_{T,i}) = \sum_{k=0}^{T-1} \Vert y_{T,i}(k \vert \cdot) - \hat{y}_{T,i}(k+1 \vert \cdot) \Vert^2$, i.e. the sum of the squared distances between steps of $y_{T,i}$ and $\hat{y}_{T,i}$, where the latter is shifted by one step.
Moreover, we use $\Vert y_{T,i} - \hat{y}_{T,i} \Vert_{T} = \sum_{k=0}^{T-1} \Vert y_{T,i}(k \vert \cdot) - \hat{y}_{T,i}(k \vert \cdot) \Vert$ to denote the sum of the distances between (not shifted) steps. 

Before we propose our distributed MPC scheme and couple agents through a suitable cost and communication, we first consider an MPC scheme for tracking a given (admissible) cooperation trajectory $r_{T,i} = (x_{T,i}, u_{T,i})$ for a single agent~\eqref{eq:agent_equations}.

Consider the stage cost
$
    \ell_i(x_i, u_i, r_{T,i}) = \Vert x_i - x_{T,i} \Vert_{Q_i}^2 + \Vert u_i - u_{T,i} \Vert_{R_i}^2
$
with $Q_i, R_i \succ 0$, and define the tracking cost
\begin{align*}
    J_i^\mr{tr}(x_i(\cdot \vert t), u_i(\cdot \vert t), r_{T,i}) = &\sum_{k=0}^{N-1} \ell_i(x_i(k \vert t), u_i(k \vert t), r_{T,i}(k \vert t))
    \\
    &{+}\: V_i^\mr{f}(x_i(N \vert t), r_{T,i}(N \vert t))
\end{align*}
with prediction horizon $N \in \mathbb{N}_0$, the $k$-step ahead predictions of the state and input trajectory $x_i(k \vert t)$ and $u_i(k \vert t)$ made at time $t$.
A suitable terminal cost $V_i^\mr{f}$ and further terminal ingredients are defined in the following assumption, adapted from \cite[Assm. 2]{Koehler2020b}.
\begin{assum}\label{asm:stabilising_terminal_ingredients}
    \begin{enumerate}
        \item There exist a terminal control law
        $
        k_{\mathrm{f},i}: \mathbb{X}_i \times \mathcal{R}_{T,i} \to \mathbb{U}_i
        $, 
        a continuous terminal cost
        $
        V_i^\mr{f}: \mathbb{X}_i \times \mathcal{R}_{T,i} \to \mathbb{R}_{\ge 0}
        $ 
        and compact terminal sets
        $
        \mathcal{X}_i^\mr{f}(r_{T,i}(k \vert t)) \subseteq {\mathbb{X}_i}
        $ 
        such that for any $r_{T,i} \in \mathcal{R}_{T,i}$ and $x_i \in \mathcal{X}_i^\mr{f}(r_{T,i}(k \vert t))$, 
        \begin{subequations}
            \begin{align}
                &V_i^\mr{f}(x_i^{+}, r_{T,i}(k+1 \vert t)) - V_i^\mr{f}(x_i, r_{T,i}(k\vert t))
                \notag
                \\
                &\le  - \ell_i(x_i, k^\mr{f}_i(x_i, r_{T,i}(k\vert t)), r_{T,i}(k\vert t)), \label{eq:terminal_cost_decrease}
                \\
                &(x_i, k^\mr{f}_i(x_i, r_{T,i}(k\vert t))) \in \mathcal{Z}_i,
                \\
                &x_i^{+} \in \mathcal{X}_i^\mr{f}(r_{T,i}(k+1\vert t)),
            \end{align}
        \end{subequations}
        for all $k \in \mathbb{I}_{0:T-1}$ where $x_i^{+} = f_i(x_i, k^\mr{f}_i(x_i, r_{T,i}(k\vert t)))$.
        \item There exist constants $c_{\mr{u},i}, \epsilon_i > 0$ for any admissible (satisfying Assumption~\ref{asm:reference}) cooperation trajectory $r_{T,i}(\cdot \vert t)$ such that for any $x_i \in \mathbb{X}_i$ with $\Vert x_i - x_{T,i}(0 \vert t) \Vert_{Q_i} \le \epsilon_i$ the MPC for tracking problem~\eqref{eq:MPC_for_tracking} is feasible and
        \begin{equation}\label{eq:tracking_value_function_upper_bound}
            W_i(x_i, r_{T,i}(\cdot \vert t)) \le c_{\mr{u},i} \Vert x_i - x_{T,i}(0 \vert t) \Vert_{Q_i}^2.
        \end{equation}
        \label{asm:stabilising_terminal_ingredients_2}
    \end{enumerate}
\end{assum}
\begin{rem}\label{rem:terminal_ingredients}
    We refer to~\cite{Koehler2020a} for the design of suitable terminal ingredients that satisfy Assumption~\ref{asm:stabilising_terminal_ingredients} given suitable assumptions on the agents~\eqref{eq:agent_equations}.
    In particular, therein, see Proposition~4 for the design of terminal equality constraints and Lemma~5 for the design of terminal costs and regions.
\end{rem}

Given the current state $x_i(t)$ and a periodic cooperation trajectory $r_{T,i}(\cdot \vert t)$, the MPC for tracking problem is
\begin{subequations}\label{eq:MPC_for_tracking}
    \begin{align}
        &W_i(x_i(t), r_{T,i}(\cdot \vert t)) = \min_{\mathclap{u_i(\cdot \vert t)}} J_i^\mr{tr}(x_i(\cdot \vert t), u_i(\cdot \vert t), r_{T,i}(\cdot \vert t))
        \label{eq:MPC_for_tracking_cost}
        \\
        \intertext{subject to}
        &x_i(0 \vert t) = x_i(t), \label{eq:MPC_for_tracking_IC}
        \\
        &x_i(k+1 \vert t) = f_i(x_i(k \vert t), u_i(k \vert t)), \; k \in \mathbb{I}_{0:N-1}, 
        \label{eq:MPC_for_tracking_dynamics}
        \\
        &(x_i(k \vert t), u_i(k \vert t)) \in \mathcal{Z}_i, k \in \mathbb{I}_{0:N-1}, 
        \label{eq:MPC_for_tracking_constraints}
        \\
        &x_i(N \vert t) \in \mathcal{X}_i^\mr{f}(r_{T,i}(N \vert t)).
        \label{eq:MPC_for_tracking_terminal}
    \end{align}
\end{subequations}

We now couple agents through a suitable cost and communication.
We start by defining the output cooperation set with an associated cooperation cost that characterises the periodic cooperative goal.
\begin{defn}[cf.~{\cite[Def. 1]{Koehler2022b}}]\label{def:cooperation_set_and_cost}
    A compact and convex set $\mathcal{Y}_T^{\mathrm{c}} \subseteq \prod_{i\in\mathcal{V}} \mathcal{Y}_{T,i}$ is called an \emph{output cooperation set} if the cooperative goal is achieved if $y_T = \mathrm{col}_{i=1}^m(y_{T, i}) \in \mathcal{Y}_T^{\mathrm{c}}$.
    A continuously differentiable function $V^\mr{c}: \prod_{i\in\mathcal{V}} \mathcal{Y}_{T,i} \to \mathbb{R}_{\ge 0}$ is an associated \emph{cooperation cost} if it has the following properties:
    \begin{enumerate}
        \item \label{en:cost_for_cooperation_measures_distance}
        There exist $\underline{\alpha}, \bar{\alpha} \in \mathcal{K}_\infty $ such that
        $
            \underline{\alpha}(\vert y_T \vert_{\mathcal{Y}^{\mathrm{c}}_T}) \le V^\mr{c} \le \bar{\alpha}(\vert y_T \vert_{\mathcal{Y}^{\mathrm{c}}_T}),
        $
        where $\vert y_T \vert_{\mathcal{Y}^{\mathrm{c}}_T} = \min_{y_T^\mathrm{c}\in \mathcal{Y}^{\mathrm{c}}_T}\Vert y_T^\mathrm{c} - y_T \Vert_{T}$.
            %%%%
        \item $V^\mr{c}$ is convex.
            %%%
        \item \label{en:cost_for_cooperation_is_separable}
        There exist $V_{ij}^\mathrm{c}: \mathcal{Y}_{T,i} \times \mathcal{Y}_{T,j} \to \mathbb{R}_{\ge 0}$ for all $(i,j) \in \mathcal{E}$ such that
        \begin{equation}\label{eq:cost_for_cooperation_distributable}
            V^\mathrm{c}(y_T) = \sum_{i\in\mathcal{V}} \sum_{j\in\mathcal{N}_i} V_{ij}^\mathrm{c}(y_{T, i}, y_{T, j}).
        \end{equation}
            %%%
        \item \label{en:cost_for_cooperation_ignores_shifting}
        For any two $y_{T, i}(\cdot \vert t) \in \mathcal{Y}_{T,i}$ and $y_{T, j}(\cdot \vert t) \in \mathcal{Y}_{T,j}$, $V_{ij}^\mathrm{c}(y_{T, i}(\cdot, t), y_{T, j}(\cdot, t)) = V_{ij}^\mathrm{c}(y_{T, i}(\cdot{+}1, t), y_{T, j}(\cdot{+}1, t))$ for all $i,j\in\mathcal{V}$.
    \end{enumerate}
\end{defn}
Property \ref{en:cost_for_cooperation_measures_distance}) means that $V^\mathrm{c}$ measures the distance of the global output to the output cooperation set, \ref{en:cost_for_cooperation_is_separable}) that $V^\mathrm{c}$ is separable according to the communication topology, and \ref{en:cost_for_cooperation_ignores_shifting}) that shifting the output cooperation trajectories does not change the cost.

Furthermore, we connect periodic cooperation output trajectories to state and input trajectories with the following assumption, similar to~\cite[Assm. 1]{Limon.2018}, \cite[Assm. 6]{Koehler2020b}, \cite[Assm. 4]{Koehler2022b}.
\begin{assum}\label{asm:output_state_input_relation}
    There exist locally Lipschitz, injective functions $g_{x,i}: \mathcal{Y}_{T,i} \to \mathbb{X}_i^T$ and $g_{u,i}: \mathcal{Y}_{T,i} \to \mathbb{U}_i^T$.
    That is,
    $x_{T, i} = g_{x,i}(y_{T, i})$ and $u_{T, i} = g_{u,i}(y_{T, i})$ are unique for any $y_{T, i} \in \mathcal{Y}_{T,i}$, and there exist $L_{g_{x,i}}, L_{g_{u,i}} > 0$ such that for any $y_{T,i}, \hat{y}_{T,i} \in \mathcal{Y}_{T,i}$ we have 
    $\Vert g_{x,i}(y_{T,i}) - g_{x,i}(\hat{y}_{T,i}) \Vert_T \le L_{g_{x,i}} \Vert y_{T,i} - \hat{y}_{T,i} \Vert_T$ and
    $\Vert g_{u,i}(y_{T,i}) - g_{u,i}(\hat{y}_{T,i}) \Vert_T \le L_{g_{u,i}} \Vert y_{T,i} - \hat{y}_{T,i} \Vert_T$.
\end{assum}
Note that we do not need to know the values of $L_{g_{x,i}}$ and $L_{g_{u,i}}$ to implement the proposed distributed MPC scheme below.

It is essential for cooperation of the agents that they also consider the influence of decisions on their (artificial) cooperation output trajectory on the costs of their neighbours.
For this purpose, we define a combined cost $\bar{V}_i^\mr{c}: \mathcal{Y}_{T,i} \times \prod_{j\in\mathcal{N}_i}\mathcal{Y}_{T,j} \to \mathbb{R}_{\ge 0}$,
\begin{equation}\label{eq:extendend_cooperation_cost}
    \bar{V}_i^\mr{c}(y_{T,i},y_{T,\mathcal{N}_i}) = \sum_{j \in \mathcal{N}_i} V_{ij}^\mr{c}(y_{T,i}, y_{T,j}) + V_{ji}^\mr{c}(y_{T,j}, y_{T,i})
\end{equation}
where $y_{T,\mathcal{N}_i}$ contains all $y_{T, j}$ with $j \in \mathcal{N}_i$. 
With this, we define the local objective function
\begin{align*}
    &J_i(x_i(t), u_i(\cdot \vert t), y_{T, i}(\cdot \vert t), \bar{y}_{T,\mathcal{N}_i}(\cdot \vert t), y_{T,i}^*(\cdot \vert t-1)) 
    \\
    &=
    J_i^\mr{tr}(x_i(\cdot \vert t), u_i(\cdot \vert t), r_{T,i}(\cdot \vert t))
    + \bar{V}_i^\mr{c}(y_{T,i}(\cdot \vert t),\bar{y}_{T,\mathcal{N}_i}(\cdot \vert t))
    \\
    &\phantom{={}} + \delta_i d(y_{T, i}(\cdot \vert t), y_{T,i}^*(\cdot \vert t-1))
\end{align*}
where $\bar{y}_{T, \mathcal{N}_i}(\cdot \vert t)$ contains the communicated (periodic) cooperation output trajectories $\bar{y}_{T, j}(\cdot \vert t)$ that have been sent to Agent $i$ by its neighbours $j\in\mathcal{N}_i$, $\delta_i>0$ is an arbitrary, small parameter, and $r_{T,i}(\cdot \vert t) = (g_{x,i}(y_{T, i}(\cdot \vert t)), g_{u, i}(y_{T, i}(\cdot \vert t)))$.
We require the term $\delta_i d$ to show in Theorem~\ref{thm:asymptotic_stability} below that the closed-loop system indeed converges to a $T$-periodic trajectory.
This is not needed in the case of cooperation at equilibria, cf.~\cite[]{Koehler2022b}.
Agent $i$'s individual MPC problem is given by
\begin{subequations}\label{eq:MPC_for_cooperation}
    \begin{align}
        &\min_{\mathclap{\substack{u_i(\cdot \vert t) \\ y_{T, i}(\cdot \vert t)}}} J_i(x_i(t), u_i(\cdot \vert t), y_{T, i}(\cdot \vert t), \bar{y}_{T, \mathcal{N}_i}(\cdot \vert t), y_{T,i}^*(\cdot \vert t-1)) 
        \label{eq:MPC_for_cooperation_objective_function}
        \\
        \intertext{subject to \eqref{eq:MPC_for_tracking_IC}, \eqref{eq:MPC_for_tracking_dynamics}, \eqref{eq:MPC_for_tracking_constraints}, \eqref{eq:MPC_for_tracking_terminal},}
        &r_{T, i}(k \vert t) = (g_{x,i}(y_{T, i}(k \vert t)), g_{u, i}(y_{T, i}(k \vert t))), k \in \mathbb{I}_{0:N},
        \\
        &y_{T, i}(\cdot \vert t) \in \mathcal{Y}_{T,i}. \label{eq:MPC_for_cooperation_cooperation_output_constraints}
    \end{align}
\end{subequations}
The solution of~\eqref{eq:MPC_for_cooperation} at time $t$ is denoted by $u_i^*(\cdot \vert t)$ and $y_{T, i}^*(\cdot \vert t)$ (if it is not unique, introduce a static mapping selecting one). 

We now state the sequential distributed MPC scheme that will lead to satisfaction of the cooperative control goal as shown in Section~\ref{sec:analysis}, which is based on~\cite[Alg. 1]{Koehler2022b}.
\begin{alg}\label{alg:sequential_MPC}
    Initialisation of Agent $i$: Omit~\eqref{eq:extendend_cooperation_cost} and the term $\delta_i d(y_{T, i}(\cdot \vert 0), y_{T,i}^*(\cdot \vert t-1))$ in~\eqref{eq:MPC_for_cooperation_objective_function}, then compute $u_i^*(\cdot \vert 0)$ and $y_{T, i}^*(\cdot \vert 0)$ by solving~\eqref{eq:MPC_for_cooperation}. Share $y_{T, i}^*(\cdot \vert 0)$ with neighbours.
    \newline
    At each time step $t \in \mathbb{N}$:
    \begin{enumerate}
        \item Sequentially, for $i = 1,\dots, m$:
        If agent $i$ has received $y_{T, j}^*(\cdot \vert t)$, set $\bar{y}_{T,j}(\cdot \vert t) = y_{T, j}^*(\cdot \vert t)$, if not, set $\bar{y}_{T,j}(\cdot \vert t) = y_{T, j}^*(\cdot{+}1 \vert t-1)$.
        Then, agent $i$ solves~\eqref{eq:MPC_for_cooperation} and sends $y_{T, i}^*(\cdot \vert t)$ to its neighbours.
        \label{alg:sequential_MPC_step1}
        \item Each agent applies $u_{\mathrm{MPC},i}(t) = u_i^*(0 \vert t)$. Go to Step 1.
    \end{enumerate}
\end{alg}
\begin{rem}
    Note that we do not need to complete a full sequence in Step~\ref{alg:sequential_MPC_step1} of Algorithm~\ref{alg:sequential_MPC} since agents may continue tracking their previously optimal trajectory using~\eqref{eq:MPC_for_tracking} as shown below in Theorem~\ref{thm:recursive_feasibility}.
    In addition, agents only have to communicate once each time step to share their optimal cooperation output trajectory.
    An interesting feature of Algorithm~\ref{alg:sequential_MPC} is that Agent $i$'s local optimisation problem~\eqref{eq:MPC_for_cooperation} is only influenced by its neighbours through~\eqref{eq:MPC_for_cooperation_objective_function}.
    Hence, agents leaving or joining the system will not stop Algorithm~\ref{alg:sequential_MPC}, and the initialisation of any agent can be completed in a decentralised way.  
\end{rem}

The global closed-loop system is given by
\begin{subequations}\label{eq:global_closed_loop}
    \begin{align}
        x(t + 1) &= f(x(t), u_\mathrm{MPC}(t)), & x(0) = x_0,
        \label{eq:global_closed_loop_a}
        \\
        y(t) &= h(x(t), u(t)), &
    \end{align}
\end{subequations}
with some initial condition $x_0 \in \mathbb{X} = \prod_{i\in\mathcal{V}} \mathbb{X}_i$, $ f = \mathrm{col}_{i\in\mathcal{V}}(f_i)$, $h = \mathrm{col}_{i\in\mathcal{V}}(h_i)$, $x(t) = \mathrm{col}_{i\in\mathcal{V}}(x_i(t))$, $u_\mathrm{MPC}(t) = \mathrm{col}_{i\in\mathcal{V}}(u_{i,\mathrm{MPC}}(t))$ and $y(t) = \mathrm{col}_{i\in\mathcal{V}}(y_i(t))$.

%===============================================================================
\section{Analysis of the closed loop}\label{sec:analysis}
In this section, we analyse properties of the sequential distributed MPC scheme in Algorithm~\ref{alg:sequential_MPC} and of the closed-loop system~\eqref{eq:global_closed_loop}.
We first show that recursive feasibility and constraint satisfaction are guaranteed for each agent individually, irrespective of the actions of other agents.
\begin{thm}\label{thm:recursive_feasibility}
    Let Assumptions~\ref{asm:reference} and \ref{asm:stabilising_terminal_ingredients} hold.
    Then, if Algorithm~\ref{alg:sequential_MPC} is applied for any initial condition $x_0$ for which~\eqref{eq:MPC_for_cooperation} is feasible, the optimisation problem~\eqref{eq:MPC_for_cooperation} in Step 1 of Algorithm~\ref{alg:sequential_MPC} is recursively feasible for all agents $i \in \mathcal{V}$.
    Consequently, the closed-loop system~\eqref{eq:global_closed_loop} satisfies the constraints, i.e. $(x(t), u_\mathrm{MPC}(t)) \in \mathcal{Z} = \prod_{i\in\mathcal{V}} \mathcal{Z}_i$ for all $t \in \mathbb{N}_0$.
\end{thm}
\begin{pf}
    Let $i \in \mathcal{V}$.
    The shifted previously optimal cooperation output trajectory is again a feasible candidate, i.e. $\hat{y}^\mr{a}_{\mr{c},i}(k \vert t+1) = y_{\mr{c},i}^*(k+1 \vert t)$, $k\in\mathbb{I}_{0:T-1}$.
    A feasible input sequence can be generated from Assumption~\ref{asm:stabilising_terminal_ingredients} by shifting the previously optimal one and appending the terminal control law, i.e. $\hat{u}^\mr{a}_i(\cdot \vert t + 1) = (u_i^*(1 \vert t), \dots, u_i^*(N-1 \vert t), k_i^\mr{f}(x_i^*(N \vert t), r_{T, i}^*(N\vert t)))$, as is standard in MPC (see, e.g.~\cite{Rawlings.2020}).
    Constraint satisfaction of the closed loop then follows from the constraints in~\eqref{eq:MPC_for_tracking_constraints} for $i \in \mathcal{V}$ with $k=0$.
    \hspace{\fill} \qed
\end{pf}

The following assumption guarantees the existence of a candidate cooperation output trajectory that is sufficiently close to the previously optimal one and in a direction that decreases the local part of the cooperation cost.
We need this candidate only for the purpose of analysing the closed loop.
Recall that $\bar{y}_{T,\mathcal{N}_i}(\cdot \vert t)$ contains the communicated cooperation output trajectories of Agent $i$'s neighbours which are fixed when Agent $i$ solves~\eqref{eq:MPC_for_cooperation}.
From Algorithm~\ref{alg:sequential_MPC}, $\bar{y}_{T,\mathcal{N}_i}(\cdot \vert t)$ contains $y_{T,j}^*(\cdot \vert t)$ for all $j\in\mathcal{N}_i$ and $j<i$, and $y_{T,j}^*(\cdot{+}1 \vert t-1)$ for all $j\in\mathcal{N}_i$ and $j>i$.
Moreover, for fixed $\bar{y}_{T,\mathcal{N}_i}(\cdot \vert t)$, the set of cooperation output trajectories $y_{T,i}$ that minimise $\bar{V}_i^\mr{c}$ is denoted by $\bar{\mathcal{Y}}_{T,i}^{\mathrm{min}}(t)$, i.e.
$
    \bar{V}_i^\mr{c}(y_{T,i}, \bar{y}_{T,\mathcal{N}_i}(\cdot \vert t)) \le \bar{V}_i^\mr{c}(\tilde{y}_{T,i} , \bar{y}_{T,\mathcal{N}_i}(\cdot \vert t))
$
holds for all $\tilde{y}_{T,i} \in \mathcal{Y}_{T,i}$ and $y_{T,i} \in \bar{\mathcal{Y}}_{T,i}^{\mathrm{min}}(t)$.
\begin{assum}\label{asm:cooperation_candidate}
    For all $i \in \mathcal{V}$ there exist $c_{1,i} > 0$, $c_{2,i} \ge 0$,
    a continuous function $\rho_{i,t}$, where $\rho_{i,t}(y_{T,i}) = 0$ if $y_{T,i} \in \bar{\mathcal{Y}}_{T,i}^{\mathrm{min}}(t)$ and $\rho_{i,t}(y_{T,i}) > 0$ otherwise, 
    and $\tilde{y}_{T,i}(\cdot\vert t+1) \in \mathcal{Y}_{T,i}$,
    with which we define for $k\in\mathbb{I}_{0:T-1}$ and $\theta_i \in (0, 1]$
    \begin{equation}\label{eq:cooperation_candidate_convex_combination}
        \hat{y}_{T,i}^{\mathrm{b}}(k \vert t+1) = (1-\theta_i)y_{T,i}^*(k+1 \vert t) + \theta_i \tilde{y}_{T,i}(k \vert t+1),
    \end{equation}
    such that for all $t\in\mathbb{N}_0$:
    \begin{align}\label{eq:candidate_norm_upper_bound}
        &\Vert y^*_{T,i}(\cdot {+} 1 \vert t) - \tilde{y}_{T,i}(\cdot \vert t + 1) \Vert_T^2 \le \rho_{i,t+1}(y_{T,i}^*(\cdot{+}1 \vert t)),
            \\%%%%%%%
        \label{eq:cooperation_cost_decrease}
        &\bar{V}_i^\mr{c}(\hat{y}_{T,i}^{\mathrm{b}}(\cdot \vert t+1)\hspace{-0.1em},\hspace{-0.1em}\bar{y}_{T,\mathcal{N}_i}(\cdot \vert t+1)) \hspace{-0.1em}-\hspace{-0.1em} \bar{V}_i^\mr{c}(y_{T,i}^*(\cdot \vert t)\hspace{-0.1em},\hspace{-0.1em}\bar{y}_{T,\mathcal{N}_i}(\cdot \vert t+1))
        \notag
            \\%%%
        &\le - (c_{1,i}\theta_i - c_{2,i}\theta_i^2) \rho_{i,t+1}(y_{T,i}^*(\cdot{+}1 \vert t)),
            \\%%%%%%
        \intertext{and}
        &\lim_{t\to\infty} \sum_{i\in\mathcal{V}} \rho_{i,t+1}(y_{T,i}^*(\cdot{+}1 \vert t)) = 0 \Longrightarrow \lim_{t\to\infty} V^\mathrm{c}(y_{T}^*(\cdot \vert t)) = 0.
        \notag
    \end{align}
\end{assum}
\begin{rem}
    Note that~\eqref{eq:candidate_norm_upper_bound} and~\eqref{eq:cooperation_cost_decrease} may be satisfied by choosing $\tilde{y}_{T,i}(k \vert t+1)$ sufficiently close to $y^*_{T,i}(k+1\vert t)$ in a direction that decreases $\bar{V}_i^\mr{c}(\tilde{y}_{T,i}(\cdot \vert t+1),\bar{y}_{T,\mathcal{N}_i}(\cdot \vert t+1))$ if $\bar{V}_i^\mr{c}( y_{T,i}^*(\cdot{+}1 \vert t),\bar{y}_{T,\mathcal{N}_i}(\cdot \vert t+1))$ is not minimal for fixed $\bar{y}_{T,\mathcal{N}_i}(\cdot \vert t+1)$.
    A possible way to satisfy Assumption~\refeq{asm:cooperation_candidate} is to assume that the gradient $\nabla_{{y}_{T,i}}\bar{V}_i^\mr{c}({y}_{T,i},\bar{y}_{T,\mathcal{N}_i})$ is Lipschitz continuous (cf.~\cite[Assm. 3]{Koehler2022b}) and to use a projected gradient-descend update with a sufficiently small step-size to show existence of $\tilde{y}_{T,i}(\cdot \vert t+1)$, cf.~\cite[Lem. 2]{Koehler2022b}.
    The term on the right-hand side of~\cite[inequality (11)]{Koehler2022b} gives the desired $\rho_{i,t+1}$ in Assumption~\ref{asm:cooperation_candidate} based on the properties of the projected gradient-descend update.
\end{rem}

Next, we show that incrementally changing the cooperation output trajectory leads to a feasible candidate in~\eqref{eq:MPC_for_cooperation} if the agent is sufficiently close to the previously optimal one, and give an upper bound on the resulting increase in the tracking cost.
\begin{lem}\label{lm:feasible_cooperation_candidate}
    Let Assumptions~\ref{asm:reference}--\ref{asm:cooperation_candidate} hold and assume~\eqref{eq:MPC_for_cooperation} is feasible at time $t$ for all $i \in \mathcal{V}$.
    Consider the following set
    \begin{equation}\label{eq:incremental_candidate_set}
        \mathcal{V}_{\mathrm{b}}
        =
        \{
            i \vert \Vert x_i(t) - x_{T, i}^*(0 \vert t) \Vert_{Q_i}^2 \le \gamma_i 
            \rho_{i,t+1}(y_{T,i}^* (\cdot{+}1 \vert t))
        \}
    \end{equation}
    with some constant $\gamma_i > 0$. % (further {restricted} below in Theorem~\ref{thm:asymptotic_stability}'s proof).
    For all $i\in\mathcal{V}_{\mathrm{b}}$ and any $\tilde{y}_{T,i}(\cdot\vert t+1) \in \mathcal{Y}_{T,i}$ there exists $\theta_i \in (0, 1]$ to determine $\hat{y}_{T,i}^{\mathrm{b}}(\cdot \vert t+1)$ through~\eqref{eq:cooperation_candidate_convex_combination} such that there exists a feasible candidate $(\hat{u}_{i}^{\mathrm{b}}(\cdot \vert t+1), \hat{y}_{T,i}^{\mathrm{b}}(\cdot \vert t+1))$ in~\eqref{eq:MPC_for_cooperation} at time $t+1$.
    Moreover, with the corresponding predicted state sequence $\hat{x}^{\mathrm{b}}_i(\cdot \vert t+1)$ and cooperation trajectory $\hat{r}_{T,i}^{\mathrm{b}}(\cdot \vert t+1)$, 
    the tracking cost $J_i^{\mathrm{tr,b}}(t+1) = J_i^{\mathrm{tr}}(\hat{x}^{\mathrm{b}}_i(\cdot \vert t+1), \hat{u}^{\mathrm{b}}_i(\cdot \vert t+1), \hat{r}^{\mathrm{b}}_{T,i}(\cdot \vert t+1))$ satisfies
    \begin{align}
        J_i^{\mathrm{tr,b}}(t+1) 
        &\le
        2c_{\mr{u},i} \Vert x_i(t+1) - x^*_{T,i}(1 \vert t) \Vert_{Q_i}^2
        \notag
            \\
        &\phantom{\le{}}
        + c_{\mathrm{tr}, i} \theta_i^2 \Vert y^*_{T,i}(\cdot{+}1 \vert t) - \tilde{y}_{T,i}(\cdot \vert t+1) \Vert_T^2
        \label{eq:candidate_tracking_cost_upper_bound}
    \end{align}
    with $c_{\mathrm{tr}, i} = 2c_{\mr{u},i} \bar{\lambda}_{Q_i} L_{g_{x,i}}^2$.
\end{lem}
\begin{pf}
    We follow similar ideas as in~\cite[proof of Thm. 8]{Koehler2020b}.
    Since $\mathcal{Y}_{T,i}$ is compact for any $i\in\mathcal{V}$, there exists $c_{\mathcal{Y}_{T,i}}>0$ such that $\rho_{i,t}(y_{T,i}) \le c_{\mathcal{Y}_{T,i}}$ for all $y_{T,i} \in \mathcal{Y}_{T,i}$ and $y_{T,\mathcal{N}_i}(\cdot\vert t) \in \prod_{j\in\mathcal{N}_i} \mathcal{Y}_{T,j}$.
    Then, if 
    $\gamma_i \le \epsilon_i^2(c_{\mathcal{Y}_{T,i}})^{-1}$ 
    with $\epsilon_i$ from Assumption~\ref{asm:stabilising_terminal_ingredients}, 
    $
        \Vert x_i(t) - x_{T,i}^*(0 \vert t) \Vert_{Q_i}^2 \stackrel{\eqref{eq:incremental_candidate_set}}{\le} \gamma_i c_{\mathcal{Y}_i} \le \epsilon_i^2
    $
    holds for all $i \in \mathcal{V}_{\mathrm{b}}$.
    From this, if also 
    $\gamma_i \le \epsilon_i^2(4 c_{\mr{u},i} c_{\mathcal{Y}_{T,i}})^{-1}$,
    \begin{align}
            &\Vert x_i^*(1 \vert t) - x_{T,i}^*(1 \vert t) \Vert_{Q_i}^2 
            \stackrel{\eqref{eq:MPC_for_tracking_cost}}{\le} W_i(x_i(t), r_{T,i}^*(\cdot \vert t))
            \notag
            \\
            &\stackrel{\eqref{eq:tracking_value_function_upper_bound},\eqref{eq:incremental_candidate_set}}{\le}
            c_{\mr{u},i} \gamma_i \rho_{i,t+1}(y_{T,i}^*(\cdot{+}1 \vert t)) \le c_{\mr{u},i} \gamma_i c_{\mathcal{Y}_{T,i}} \le \frac{1}{4}\epsilon_i^2.
            \label{eq:bound_future_target_distance}
    \end{align}
    From Assumption~\ref{asm:output_state_input_relation}, there exists a cooperation state trajectory corresponding to the cooperation output trajectory, 
    $\hat{x}^\mr{b}_{T,i}(\cdot \vert t+1) = g_{x,i}(\hat{y}^\mr{b}_{T,i}(\cdot \vert t+1))$.
    Then, for all $i \in \mathcal{V}_\mr{b}$, since $x(t+1) = x^*(1 \vert t)$ by~\eqref{eq:global_closed_loop_a} and if $\theta_i \le \epsilon_i(2L_{g_{x,i}} \sqrt{\bar{\lambda}_{Q_i}} T c_{\mathcal{Y}_i})^{-1}$
    \begin{align*}
        &\Vert x_i(t+1) - \hat{x}^{\mr{b}}_{T, i}(0 \vert t+1) \Vert_{Q_i} 
            \\%%%
        &\le \Vert x_i(t+1) - x_{T, i}^*(1 \vert t) \Vert_{Q_i} 
            \\
        &\phantom{\le{}} + \sqrt{\bar{\lambda}_{Q_i}}
        \Vert x_{T, i}^*(\cdot{+}1 \vert t) - \hat{x}^{\mathrm{b}}_{T, i}(\cdot \vert t+1) \Vert_T
        \\%%%
        &\stackrel{\mathclap{\text{Assm.~\ref{asm:output_state_input_relation}},\,\eqref{eq:cooperation_candidate_convex_combination}}}{\le} \Vert x_i(t+1) - x_{T, i}^*(1 \vert t) \Vert_{Q_i} 
        \\
        &\phantom{= {}} + L_{g_{x,i}} \sqrt{\bar{\lambda}_{Q_i}}
        \theta_i \Vert y_{T,i}^*(\cdot{+}1 \vert t) - \tilde{y}_{T,i}(\cdot \vert t+1) \Vert_T
        \\%%%
        &\le \Vert x_i(t+1) - x_{T, i}^*(1 \vert t) \Vert_{Q_i} + L_{g_{x,i}} \sqrt{\bar{\lambda}_{Q_i}} T \theta_i c_{\mathcal{Y}_i} \stackrel{\eqref{eq:bound_future_target_distance}}{\le} \epsilon_i
    \end{align*}
    where the constant $c_{\mathcal{Y}_i} \ge \Vert y_{T,i}^*(k+1 \vert t) - \tilde{y}_{T,i}(k \vert t+1) \Vert$, $k \in \mathbb{I}_{0:T-1}$, exists by compactness of $\mathcal{Y}_{T,i}$.
    Hence, Assumption~\ref{asm:stabilising_terminal_ingredients} ensures that a feasible candidate input sequence $\hat{u}^\mr{b}_i(\cdot \vert t+1)$ exists such that $\hat{u}^\mr{b}_i(\cdot \vert t+1)$ and $\hat{y}_{T,i}^{\mathrm{b}}(\cdot \vert t+1)$ are feasible candidates in~\eqref{eq:MPC_for_cooperation}.
    Furthermore, also from Assumption~\ref{asm:stabilising_terminal_ingredients},
    \begin{align*}
        &J_i^\mr{tr}(\hat{x}^\mr{b}_i(\cdot \vert t+1), \hat{u}^\mr{b}_i(\cdot \vert t+1), \hat{r}^\mr{b}_{T,i}(\cdot \vert t+1))
        \notag
        \\
        &\stackrel{\eqref{eq:tracking_value_function_upper_bound}}{\le}
        c_{\mr{u},i} \Vert x_i(t+1) - \hat{x}^\mr{b}_{T,i}(0 \vert t+1) \Vert_{Q_i}^2 
        \notag
        \\%%%
        &\le 2c_{\mr{u},i} \Vert x_i(t+1) - x^*_{T,i}(1\vert t) \Vert_{Q_i}^2 
        \\
        &\phantom{\le{}} + 2c_{\mr{u},i} \bar{\lambda}_{Q_i} 
        \Vert x^*_{T,i}(\cdot{+}1\vert t) - \hat{x}^\mr{b}_{T,i}(\cdot \vert t+1) \Vert_T^2
        \\%%%
        &\stackrel{\mathclap{\text{Assm.~\ref{asm:output_state_input_relation}},\, \eqref{eq:cooperation_candidate_convex_combination}}}{\le} 2c_{\mr{u},i} \Vert x_i(t+1) - x^*_{T,i}(1\vert t) \Vert_{Q_i}^2 
        \\
        &\phantom{\le{}} + 2c_{\mr{u},i} \bar{\lambda}_{Q_i} L_{g_{x,i}}^2 \theta_i^2
        \Vert y^*_{T,i}(\cdot{+}1\vert t) - \tilde{y}_{T,i}(\cdot \vert t+1) \Vert_T^2
    \end{align*}
    where $\hat{x}^\mr{b}_i(\cdot \vert t+1)$ is the state trajectory that results from applying $\hat{u}^\mr{b}_i(\cdot \vert t+1)$ starting at $x_i(t+1)$, showing~\eqref{eq:candidate_tracking_cost_upper_bound}.
    \hspace{\fill} \qed
\end{pf}

To prove convergence of the closed-loop system, we define
$
    V(t) 
    = 
    V^\mr{c}(y_{T}^*(\cdot \vert t))
    +
    \sum_{i\in\mathcal{V}}
    (J_i^\mr{tr}(x_i^*(\cdot \vert t), u_i^*(\cdot \vert t), r_{T,i}^*(\cdot \vert t))
    +
    \delta_i d(y_{T,i}^*(\cdot \vert t), y_{T,i}^*(\cdot \vert t-1))),
$
and show in the following that $V(t)$, which is bounded from below, is non-increasing and therefore converges.
We start by providing an upper bound on the difference between two consecutive time steps.

\begin{lem}\label{lm:value_function_upper_bound}
    Let Assumptions~\ref{asm:reference}--\ref{asm:cooperation_candidate} hold and let~\eqref{eq:MPC_for_cooperation} be feasible at time $t$.
    Define
    \begin{equation}\label{eq:standard_candidate_set}
        \mathcal{V}_\mr{a}
        =
        \{
            i \vert \Vert x_i(t) - x_{T, i}^*(0 \vert t) \Vert_{Q_i}^2 \hspace{-0.1em}\ge\hspace{-0.1em} \gamma_i \rho_{i,t+1}(y_{T,i}^* (\cdot{+}1 \vert t))
        \}
    \end{equation}
    and consider again the set $\mathcal{V}_\mr{b}$ from~\eqref{eq:incremental_candidate_set}.
    Take $(\hat{u}_i^\mr{a}(\cdot \vert t + 1), \hat{y}_{T,i}^\mr{a}(\cdot \vert t+1))$ from Theorem~\ref{thm:recursive_feasibility} and $(\hat{u}_i^\mr{b}(\cdot \vert t + 1), \hat{y}_{T,i}^\mr{b}(\cdot \vert t+1))$ from Lemma~\ref{lm:feasible_cooperation_candidate}.
    Define the shorthands
    \begin{align*}
        {J}_i^{\mr{tr},*}(t) &=  J_i^{\mr{tr}}(x^*_i(\cdot \vert t), u^*_i(\cdot \vert t), r^*_{T,i}(\cdot \vert t)),
        \\
        \hat{J}_i^{\mr{tr,b}}(t+1) &=  J_i^\mr{tr}(\hat{x}^\mr{b}_i(\cdot \vert t+1), \hat{u}^\mr{b}_i(\cdot \vert t+1), \hat{r}^\mr{b}_{T,i}(\cdot \vert t+1)).
    \end{align*}
    Then,
    \begin{align}\label{eq:candidate_upper_bound}
        &V(t+1) - V(t)
        \notag
        \\
        &\le - \sum_{i \in \mathcal{V}_\mr{b} } {J}_i^{\mr{tr},*}(t) - \sum_{i \in \mathcal{V}} \delta_i d(y_{T,i}^*(\cdot \vert t), y_{T,i}^*(\cdot \vert t-1))
        \notag
        \\
        &\phantom{={}}- \sum_{i \in \mathcal{V}_\mr{a} } \ell_i(x_i(t), u_i^*(0 \vert t), r_{T,i}^*(0 \vert t))
        \notag
        \\
        &\phantom{={}} + \sum_{i \in \mathcal{V}_\mr{b} } \hat{J}_i^\mr{tr,b}(t+1)
        + \delta_i d(y_{T,i}^{\mathrm{b}}(\cdot \vert t+1), y_{T,i}^*(\cdot \vert t))
        \notag
        \\
        &\phantom{\le{}} - \sum_{i \in \mathcal{V}_\mr{b} } (c_{1,i}\theta_i - c_{2,i}\theta_i^2) \rho_{i,t+1}(y_{T,i}^*(\cdot{+}1 \vert t)) .
    \end{align}
\end{lem}
\begin{pf}
    We define
    $
    z^i_{t+1} = (y_{T, 1}^*(\cdot \vert t+1), \dots, y_{T, i-1}^*(\cdot \vert t+1), y_{T, i}^*(\cdot{+}1 \vert t), \dots, y^*_{T, m}(\cdot{+}1 \vert t))
    $,
    containing the cooperative output trajectories at the beginning of the $i$-th step of the sequence in Step 1 of Algorithm~\ref{alg:sequential_MPC}.
    That is, Agents~1 to $i-1$ have updated their cooperative output, whereas Agents~$i$ to $m$ have yet to optimise, but their communicated previously optimal cooperation output trajectories from time $t-1$ are available.
    In addition, $z^{m+1}_{t+1} = (y_{T, 1}^*(\cdot \vert t+1), \dots, y^*_{T, m}(\cdot \vert t+1))$.
    With a convenient abuse of notation, we write $V_{ij}^\mr{c}(z^i_{t+1})$, which, in fact, only depends on the $i$-th and $j$-th components of $z^i_{t+1}$.
    Since a feasible candidate is an upper bound on the optimal cost, both candidates $(\hat{u}_i^\mathrm{a}(\cdot \vert t+1), \hat{y}_{T,i}^\mathrm{a}(\cdot \vert t+1))$ and $(\hat{u}_i^\mathrm{b}(\cdot \vert t+1), \hat{y}_{T,i}^\mathrm{b}(\cdot \vert t+1))$ provide an upper bound on the optimal cost of Agent~$i$ in~\eqref{eq:MPC_for_cooperation}. 
    We use the additional shorthands
    \begin{align*}
        \bar{J}_i^{\mr{tr,a}}(t+1) &= {J}_i^{\mr{tr},*}(t) - \ell_i(x_i(t), u_i^*(0 \vert t), r_{T,i}^*(0 \vert t)),
        \\
        \bar{J}_i^{\mr{tr,b}}(t+1) &= \hat{J}_i^{\mr{tr,b}}(t+1) + \delta_i d(\hat{y}_{T,i}^{\mathrm{b}}(\cdot \vert t+1), y_{T,i}^*(\cdot \vert t))
        \\
        &\phantom{={}} - (c_{1,i}\theta_i - c_{2,i}\theta_i^2) \rho_{i,t+1}(y_{T,i}^*(\cdot{+}1\vert t)).
    \end{align*}
    If $(\hat{u}_i^\mathrm{a}(\cdot \vert t+1), \hat{y}_{T,i}^\mathrm{a}(\cdot \vert t+1))$ is used, the upper bound is
    $
        J_i^{\mr{tr},*}(t+1) + \sum_{j \in \mathcal{N}_i} \left( V_{ij}^\mr{c}(z_{t+1}^{i+1}) + V_{ji}^\mr{c}(z_{t+1}^{i+1}) \right) 
        + \delta_i d(y_{T, i}^*(\cdot \vert t+1), y_{T,i}^*(\cdot \vert t))
        \le \bar{J}_i^{\mr{tr,a}}(t+1) + \sum_{j \in \mathcal{N}_i} \left(V_{ij}^\mr{c}(z_{t+1}^{i}) + V_{ji}^\mr{c}(z_{t+1}^{i})\right)
    $
    where the inequality follows from standard arguments in MPC (cf.~\cite{Rawlings.2020}).
    If $(\hat{u}_i^\mathrm{b}(\cdot \vert t+1), \hat{y}_{T,i}^\mathrm{b}(\cdot\vert t+1))$ is used, we have 
    $
        J_i^{\mr{tr},*}(t+1) + \sum_{j \in \mathcal{N}_i} \left( V_{ij}^\mr{c}(z_{t+1}^{i+1}) + V_{ji}^\mr{c}(z_{t+1}^{i+1}) \right)
        + \delta_i d(y_{T, i}^*(\cdot \vert t+1), y_{T,i}^*(\cdot \vert t))
        \stackrel{\eqref{eq:cooperation_cost_decrease}}{\le} 
        \bar{J}_i^{\mr{tr,b}}(t+1) 
        + \sum_{j \in \mathcal{N}_i} (V_{ij}^\mr{c}(z_{t+1}^{i}) + V_{ji}^\mr{c}(z_{t+1}^{i}) ).
    $
    From here, it is straightforward to follow the steps in~\cite[proof of Lem. 5]{Koehler2022b} to arrive at
    $
        V(t+1) \le \sum_{i\in\mathcal{V}} \bar{J}_i^{\mr{tr}}(t+1) 
        + \sum_{j \in \mathcal{N}_i} V_{ij}^\mr{c}(z_{t+1}^{1})
    $
    where we have $V_{ij}^\mr{c}(z_{t+1}^{1}) = V_{ij}^\mr{c}(y_{T, i}^*(\cdot \vert t), y_{T, j}^*(\cdot \vert t))$ for all $i,j \in \mathcal{V}$.
    Inserting $\bar{J}_i^{\mr{tr}}(t+1) = \bar{J}_i^{\mr{tr,a}}(t+1)$ if $i \in \mathcal{V}_\mr{a}$ or $\bar{J}_i^{\mr{tr}}(t+1) = \bar{J}_i^{\mr{tr,b}}(t+1)$ if $i \in \mathcal{V}_\mr{b}$, and adding and subtracting $V(t)$ yields~\eqref{eq:candidate_upper_bound}. \hspace{\fill} \qed
\end{pf}

Finally, we prove that the closed loop asymptotically converges to a solution of the periodic dynamic cooperative control goal.
\begin{thm}\label{thm:asymptotic_stability}
    Let Assumptions~\ref{asm:reference}--\ref{asm:cooperation_candidate} hold.
    Then, if Algorithm~\ref{alg:sequential_MPC} is applied for any $x_0$ for which its initialisation is feasible, the closed-loop system~\eqref{eq:global_closed_loop} converges to a periodic state trajectory
    with a corresponding output trajectory that solves the cooperative goal.
\end{thm}
\begin{pf}
    Recall~\eqref{eq:candidate_upper_bound}, where we neglect some non-positive terms:
    \begin{align}\label{eq:Lyapunov_decrease_upper_bound_1}
        &V(t+1) - V(t)
        \notag\\
        &\le - \sum_{i \in \mathcal{V}} \Vert x_i(t) - x_{T,i}^*(0 \vert t) \Vert_{Q_i}^2 
        - \delta_i d(y_{T,i}^*(\cdot \vert t), y_{T,i}^*(\cdot \vert t-1))
        \notag\\
        &\phantom{\le} + \sum_{i \in \mathcal{V}_\mr{b} } \big( \hat{J}_i^\mr{tr,b}(t+1) - (c_{1,i}\theta_i - c_{2,i}\theta_i^2) \rho_{i,t+1}(y_{T,i}^*(\cdot{+}1\vert t)) \big)
        \notag
        \\
        &\phantom{=} + \sum_{i \in \mathcal{V}_{\mathrm{b}}} \delta_i d(y_{T,i}^\mathrm{b}(\cdot \vert t+1), y_{T,i}^*(\cdot \vert t))
        %\sum_{k=0}^{T-1} \Vert y_{T,i}^\mathrm{b}(k \vert t+1) - y_{T,i}^*(k+1 \vert t) \Vert^2
    \end{align}
    Note that~\eqref{eq:candidate_norm_upper_bound} implies $d(\tilde{y}_{T,i}(k \vert t+1), y^*_{T,i}(k+1\vert t)) \le \rho_{i,t+1}(y_{T,i}^*(\cdot{+}1 \vert t))$.
    Hence,
    \begin{align*}
        &V(t+1) - V(t) 
        \\
        &\stackrel{\quad\mathclap{\eqref{eq:cooperation_candidate_convex_combination},\eqref{eq:candidate_tracking_cost_upper_bound}}}{\le} 
        - \sum_{i \in \mathcal{V}} \Vert x_i(t) - x_{T,i}^*(0 \vert t) \Vert_{Q_i}^2 - \delta_i 
        d(y_{T,i}^*(\cdot \vert t), y_{T,i}^*(\cdot \vert t-1))
        \\
        &\phantom{\le{}} - \sum_{i \in \mathcal{V}_\mr{b} } (c_{1,i}\theta_i - c_{2,i}\theta_i^2) \rho_{i,t+1}(y_{T,i}^*(\cdot{+}1\vert t))
        \\
        &\phantom{\le{}} + \sum_{i \in \mathcal{V}_\mr{b} }  2c_{\mr{u},i} \Vert x_i(t+1) - x^*_{T,i}(1\vert t) \Vert_{Q_i}^2
        \\
        &\phantom{\le{}} + \sum_{i \in \mathcal{V}_\mr{b} } c_{\mathrm{tr},i} \theta_i^2 
        \Vert y_{T,i}^*(\cdot{+}1 \vert t) - \tilde{y}_{T,i}(\cdot \vert t+1) \Vert_T^2
        \\
        &\phantom{={}} + \sum_{i \in \mathcal{V}_{\mathrm{b}}} \theta_i^2 \delta_i 
        d(\tilde{y}_{T,i}(\cdot \vert t+1), y_{T,i}^*(\cdot \vert t))
        \notag
            \\%%%
        &\stackrel{\eqref{eq:incremental_candidate_set},\eqref{eq:bound_future_target_distance},\eqref{eq:candidate_norm_upper_bound}}{\le}
        - \sum_{i \in \mathcal{V}} \Vert x_i(t) - x_{T,i}^*(0 \vert t) \Vert_{Q_i}^2
        \\
        &\phantom{={}} - \sum_{i \in \mathcal{V}} \delta_i d(y_{T,i}^*(\cdot \vert t), y_{T,i}^*(\cdot \vert t-1))
        \\
        &\phantom{={}} + \sum_{i \in \mathcal{V}_\mr{b} } (2c_{\mr{u},i}^2\gamma_i - \sigma_i) \rho_{i,t+1}(y_{T,i}^*(\cdot{+}1\vert t))
    \end{align*}
    where $\sigma_i = ((c_{\mathrm{tr},i} + c_{2,i} + \delta_i )\theta_i - c_{\mathrm{1},i})\theta_i$ is positive if $\theta_i < c_{\mathrm{1},i}(c_{\mathrm{tr},i} + c_{2,i} + \delta_i )^{-1}$.
    If, also, $\gamma_i \le \sigma_i(4c_{\mathrm{u},i}^2)^{-1}$, then
    $
        V(t+1) - V(t)
        \le - {\sum_{i \in \mathcal{V}}} \frac{1}{2}\Vert x_i(t) - x_{T,i}^*(0 \vert t) \Vert_{Q_i}^2 - {\sum_{i \in \mathcal{V}_\mr{b}}} \frac{\sigma_i}{2} \rho_{i,t+1}(y_{T,i}^*(\cdot{+}1\vert t))
        - {\sum_{i \in \mathcal{V}_{\mathrm{a}}}} \frac{1}{2}\Vert x_i(t) - x_{T,i}^*(0 \vert t) \Vert_{Q_i}^2
        - \sum_{i \in \mathcal{V}} \delta_i d(y_{T,i}^*(\cdot \vert t), y_{T,i}^*(\cdot \vert t-1))
    $.
    Finally, from~\eqref{eq:standard_candidate_set} and with $\mu = \min_{i\in\mathcal{V}}\{\sigma_i, \gamma_i\}$, we get
    \begin{align}\label{eq:Lyapunov_function_decrease}
        &V(t+1) - V(t) 
        \notag
            \\%%%
        &\le - \frac{1}{2}\sum_{i \in \mathcal{V}} \left( \Vert x_i(t) - x_{T,i}^*(0 \vert t) \Vert_{Q_i}^2 + \mu \rho_{i,t+1}(y_{T,i}^*(\cdot{+}1\vert t)) \right)
        \notag
        \\
        &\phantom{={}} - \sum_{i \in \mathcal{V}} \delta_i \sum_{k=0}^{T-1} \Vert y_{T,i}^*(k \vert t) - y_{T,i}^*(k+1 \vert t-1) \Vert^2.
    \end{align}
    Since $V(t)$ is non-increasing and bounded from below, it converges.
    Hence, for all $i \in \mathcal{V}$, $\lim_{t\to\infty} y_{T,i}^*(k \vert t) - y_{T,i}^*(k+1 \vert t-1) = 0$, which holds for all $k \in \mathbb{N}_0$ since $y_{T,i}^*(\cdot \vert t)$ is $T$-periodic.
    Thus, for all $k \in \mathbb{I}_{0:T-1}$, $y_{T,i}^*(k \vert t) \to y_{T,i}^*(0 \vert t-T+k)$ as $t \to \infty$.
    Let $k\in \mathbb{I}_{T:2T-1}$ and define $\kappa = k - T$, then $y_{T,i}^*(\kappa \vert t) = y_{T,i}^*(\kappa+T \vert t) \to y_{T,i}^*(0 \vert t+\kappa)$ as $t\to\infty$ for all $\kappa \in \mathbb{I}_{0:T-1}$.
    Moreover, $\lim_{t \to \infty} \sum_{i\in\mathcal{V}} \rho_{i,t+1}(y_{T,i}^*(\cdot{+}1 \vert t)) = 0$ for all $i \in \mathcal{V}$,
    which implies $y_{T,i}^*(\cdot \vert t) \to \mathcal{Y}_{T}^{\mathrm{c}}$ as $t \to \infty$ by Assumption~\ref{asm:cooperation_candidate} and Definition~\ref{def:cooperation_set_and_cost}.
    Since $g_{x,i}$ is continuous for all $i\in\mathcal{V}$, as $t \to \infty$, $x_{T}^*(k \vert t) \to g_{x}(y_{T}^*(k \vert t))$.
    In addition, from~\eqref{eq:Lyapunov_function_decrease} it follows that $x_i(t) \to x_{T,i}^*(0 \vert t)$ as $t\to\infty$.
    Thus, for all $\kappa \in \mathbb{I}_{0:T-1}$, we have $x_i(t+\kappa) \to g_{x,i}(y_{T,i}^*(0 \vert t+\kappa)) \to g_{x,i}(y_{T,i}^*(\kappa \vert t))$ for all $i\in\mathcal{V}$ as $t\to\infty$.
    Hence, the closed-loop converges to a periodic state trajectory that corresponds to an output trajectory which achieves the cooperative goal.
    \hspace{\fill} \qed
\end{pf}
%

%===============================================================================
\section{Example: Synchronisation}
We illustrate the application of the framework to the task of synchronisation.
That is, we want the agents to converge to a common periodic trajectory.
Consider $m=4$ agents with all-to-all communication and double-integrator dynamics
\begin{align*}
    x_i(t+1) &= \begin{bmatrix} 1 & 0 & 1 & 0 \\ 0 & 1 & 0 & 1 \\ 0 & 0 & 1 & 0 \\ 0 & 0 & 0 & 1 \end{bmatrix} x_i(t) + \begin{bmatrix} 0 & 0 \\ 0 & 0 \\ 1 & 0\\ 0 & 1 \end{bmatrix} u_i(t), 
    &x_i(0) = x_i^\mr{ic},
    \\
    y_i(t) &= x_i(t), &
\end{align*}
with $x_1^\mr{ic} = \begin{bmatrix} 1.5 & 0.9 & 0 & 0 \end{bmatrix}^\top$, $x_2^\mr{ic} = \begin{bmatrix} 1 & 2 & 0 & 0 \end{bmatrix}^\top$, $x_3^\mr{ic} = \begin{bmatrix} 1.5 & 2 & 0 & 0 \end{bmatrix}^\top$, $x_4^\mr{ic} = \begin{bmatrix} 1.4 & 1.35 & 0 & 0 \end{bmatrix}^\top$,
and constraints
$
    {\mathbb{X}_i} = \{ x_i \in \mathbb{R}^4 \mid \Vert \begin{bmatrix} x_{i,1} \\ x_{i,2} \end{bmatrix}  \Vert_\infty \le 4.1, \; \Vert \begin{bmatrix} x_{i,3} \\ x_{i,4} \end{bmatrix} \Vert_\infty \le 2.1 \},
$ 
as well as
$
    \mathbb{U}_i = \{ u_i \in \mathbb{R}^2 \mid \Vert u_i \Vert_\infty \le 1.1 \} 
$ 
for all $i \in \mathcal{V}$.
We allow all periodic cooperation output trajectories with $T=10$ that satisfy tightened constraints, i.e.
$
    \mathcal{Y}_{T,i} = \{ r_{T,i} = (x_{T,i}, u_{T,i}) \in \bar{\mathbb{X}}_{i}^T \times \bar{\mathbb{U}}_{i}^T \mid
    r_{T,i} \text{ follows the dynamics above and is $T$-periodic} \}
$
with 
$
    \bar{\mathbb{X}}_i = \{ x_i \in \mathbb{R}^4 \mid \Vert \begin{bmatrix} x_{i,1} \\ x_{i,2} \end{bmatrix}  \Vert_\infty \le 4, \; \Vert \begin{bmatrix} x_{i,3} \\ x_{i,4} \end{bmatrix} \Vert_\infty \le 2 \},
$
and $\bar{\mathbb{U}}_i = \{ u_i \in \mathbb{R}^2 \mid \Vert u_i \Vert_\infty \le 1 \}$.
We choose $\delta_i=10^{-7}$ for all $i\in\mathcal{V}$ and implement Algorithm~\ref{alg:sequential_MPC} with terminal equality constraints (cf. Remark~\ref{rem:terminal_ingredients}) and a horizon of $N=10$
using CasADi~\cite[]{Andersson2019} and Ipopt~\cite[]{Waechter2005}. All numerical simulations were implemented in Python.
Finally, since we want to achieve synchronisation to a periodic trajectory, we choose $V^\mathrm{c}(y_T) = \sum_{i\in\mathcal{V}}\sum_{j\in\mathcal{N}_j} \sum_{k=0}^{T-1} \Vert y_{T,i}(k\vert t) - y_{T,j}(k\vert t) \Vert^2$.
The simulation results are depicted in Figures~\ref{fig:synchronisation_time_evolution} and~\ref{fig:synchronisation_positions}, which show that the control goal is achieved.
\begin{figure}[t]
    \setlength\axisheight{0.6\linewidth}
    \setlength\axiswidth{0.95\linewidth}
    \centering
    \begin{tikzpicture}

\definecolor{red}{RGB}{214,39,40}
\definecolor{grey}{RGB}{176,176,176}
\definecolor{orange}{RGB}{255,127,14}
\definecolor{green}{RGB}{44,160,44}
\definecolor{lgrey}{RGB}{204,204,204}
\definecolor{blue}{RGB}{31,119,180}

\begin{axis}[
height=\axisheight,
legend cell align={left},
legend style={fill opacity=0.8, 
              draw opacity=1,
              text opacity=1,
              anchor=south west,
              at={(0.65,0.03)},
              draw=lgrey,
              font=\small},
minor tick num=2,
minor xtick={-1,1,2,3,4,6,7,8,9,11,12,13,14,16,17,18,19,21,22,23,24,26,27,28,29,31},
minor ytick={0.98,1.02,1.04,1.06,1.08,1.12,1.14,1.16,1.18,1.22,1.24,1.26,1.28,1.32,1.34,1.36,1.38,1.42,1.44,1.46,1.48,1.52},
tick align=outside,
tick pos=left,
width=\axiswidth,
x grid style={grey},
xlabel={\(\displaystyle t\) (time steps)},
xmajorgrids,
xmin=-1.5, xmax=31.5,
xtick distance=10,
xtick style={color=black},
xtick={-5,0,5,10,15,20,25,30,35},
y grid style={grey},
ylabel={\(\displaystyle y_{i,1}\)},
ymajorgrids,
ymin=0.975, ymax=1.525,
ytick distance=0.2,
ytick style={color=black},
ytick={0.9,1,1.1,1.2,1.3,1.4,1.5,1.6},
label style={font=\small},
tick label style={font=\footnotesize}
]
\addplot [blue, thick]
table {%
0 1.5
1 1.5
2 1.36367189822265
3 1.29795738479325
4 1.2688497288823
5 1.26372354981297
6 1.27524397255271
7 1.29997214150768
8 1.33396177470427
9 1.3659068303146
10 1.36484437975594
11 1.34537897087761
12 1.30373658852882
13 1.27606165067014
14 1.26235388022253
15 1.26228168038501
16 1.2751083121511
17 1.30003137986278
18 1.33397212723407
19 1.36587317749519
20 1.36485843959436
21 1.34542182196256
22 1.30383006625521
23 1.2761250538856
24 1.26239022479316
25 1.26228855723797
26 1.275084694485
27 1.29997747916682
28 1.33389410239472
29 1.36579616961187
30 1.36485529058592
};
\addlegendentry{Agent 1}
\addplot [orange, dotted, thick]
table {%
0 1
1 1
2 1.13476055414273
3 1.2085497755265
4 1.23983598329631
5 1.25636500571885
6 1.27422627403073
7 1.30028739681784
8 1.33431249261235
9 1.36610037780524
10 1.36492681952048
11 1.34540651673254
12 1.30374322638613
13 1.27606146151523
14 1.26235375858237
15 1.26228230665564
16 1.27510935366353
17 1.30003253726
18 1.33397325153109
19 1.36587409882271
20 1.3648586875966
21 1.34542019000646
22 1.30382821223971
23 1.27612310580354
24 1.26239006550295
25 1.26228939669277
26 1.27508588433952
27 1.29997870744922
28 1.33389525434277
29 1.36579710029422
30 1.36485554238365
};
\addlegendentry{Agent 2}
\addplot [green, dash pattern=on 1pt off 3pt on 3pt off 3pt, thick]
table {%
0 1.5
1 1.5
2 1.39206750477056
3 1.3139033504146
4 1.27588477550655
5 1.26618547020561
6 1.2758866760767
7 1.30004756305792
8 1.33391764916176
9 1.36586518108483
10 1.36482141367878
11 1.34536546248274
12 1.3037291241081
13 1.27605644931964
14 1.26235344208633
15 1.26228359305843
16 1.27511098123899
17 1.30003411398415
18 1.33397469549538
19 1.365875269818
20 1.36485902976331
21 1.34541820130289
22 1.30382596523038
23 1.27612072025615
24 1.26238987907946
25 1.26229040615731
26 1.27508730510639
27 1.29998018191174
28 1.33389665753283
29 1.36579825817412
30 1.36485588281773
};
\addlegendentry{Agent 3}
\addplot [red, dashed, thick]
table {%
0 1.4
1 1.4
2 1.33136863731788
3 1.28768234739805
4 1.26634412070199
5 1.26339464420366
6 1.27534025539984
7 1.30007565542465
8 1.33401763345289
9 1.36593106879197
10 1.36485232200044
11 1.34537479941008
12 1.30373014102333
13 1.27605441130877
14 1.26235330311073
15 1.26228473286411
16 1.27511261748591
17 1.30003585464789
18 1.33397639460846
19 1.36587670649909
20 1.36485948138612
21 1.34541577152983
22 1.30382323805451
23 1.27611779467721
24 1.26238966122165
25 1.26229162122154
26 1.27508900491434
27 1.29998195566932
28 1.33389837038618
29 1.36579970004425
30 1.36485633790141
};
\addlegendentry{Agent 4}
\end{axis}

\end{tikzpicture}
    \caption{Time evolution of the first output of all agents.}
    \label{fig:synchronisation_time_evolution}
\end{figure}
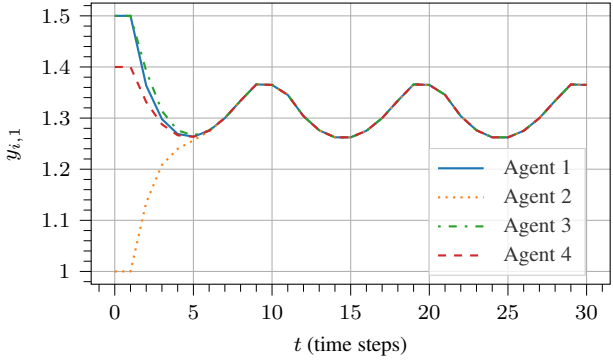
\begin{figure}[t]
    \setlength\axisheight{0.65\linewidth}
    \setlength\axiswidth{0.95\linewidth}
    \centering
    \begin{tikzpicture}

\definecolor{red}{RGB}{214,39,40}
\definecolor{grey}{RGB}{176,176,176}
\definecolor{orange}{RGB}{255,127,14}
\definecolor{green}{RGB}{44,160,44}
\definecolor{lgrey}{RGB}{204,204,204}
\definecolor{blue}{RGB}{31,119,180}

\begin{axis}[
height=\axisheight,
legend cell align={left},
legend style={
  fill opacity=0.8,
  draw opacity=1,
  text opacity=1,
  at={(0.03,0.03)},
  anchor=south west,
  draw=lgrey,
  font=\small
},
minor tick num=2,
minor xtick={},
minor ytick={},
tick align=outside,
tick pos=left,
width=\axiswidth,
x grid style={grey},
xlabel={\(\displaystyle y_{i,1}\)},
xmajorgrids,
xmin=0.975, xmax=1.525,
xtick distance=10,
xtick style={color=black},
xtick={0.9,1,1.1,1.2,1.3,1.4,1.5,1.6},
y grid style={grey},
ylabel={\(\displaystyle y_{i,2}\)},
ymajorgrids,
ymin=0.845, ymax=2.055,
ytick distance=0.2,
ytick style={color=black},
label style={font=\small},
tick label style={font=\footnotesize},
ytick={0.8,1,1.2,1.4,1.6,1.8,2,2.2}
]
\addplot [blue, thick]
table {%
1.5 0.9
1.5 0.9
1.36367189822265 1.24730927671075
1.29795738479325 1.42809432874941
1.2688497288823 1.50160466728605
1.26372354981297 1.534096229468
1.27524397255271 1.55923151071189
1.29997214150768 1.59107614333468
1.33396177470427 1.63057434182564
1.3659068303146 1.66329906945715
1.36484437975594 1.64442052092345
1.34537897087761 1.60518184472892
1.30373658852882 1.58334531213049
1.27606165067014 1.55782475154538
1.26235388022253 1.54395442912585
1.26228168038501 1.54490856744823
1.2751083121511 1.56074964127108
1.30003137986278 1.5905898820421
1.33397212723407 1.62999122716709
1.36587317749519 1.66294517208033
1.36485843959436 1.644340555951
1.34542182196256 1.60522271124353
1.30383006625521 1.58337928544833
1.2761250538856 1.55788425948793
1.26239022479316 1.54399138502301
1.26228855723797 1.54491282003364
1.275084694485 1.56071888466047
1.29997747916682 1.59052437766909
1.33389410239472 1.62989992889791
1.36579616961187 1.66286453550599
1.36485529058592 1.64437734912815
};
\addlegendentry{Agent 1}
\addplot [line width=0.04pt, black, mark=x, mark size=1.5, mark options={solid}, forget plot, thick]
table {%
1.36488392 1.64443379
1.34461458 1.6042903
1.30176498 1.58080626
1.27345644 1.55465682
1.25950324 1.54053725
1.25921113 1.54123424
1.27182452 1.55680029
1.29687909 1.58676614
1.33183833 1.62737731
1.36571511 1.66277172
1.36488392 1.64443379
};
\addplot [thick, orange, dotted, thick]
table {%
1 2
1 2
1.13476055414273 1.81622789585227
1.2085497755265 1.65680854176638
1.23983598329631 1.57907820245416
1.25636500571885 1.55479925695032
1.27422627403073 1.5625343281175
1.30028739681784 1.59044067782626
1.33431249261235 1.6296826445205
1.36610037780524 1.6627812368581
1.36492681952048 1.64419184700259
1.34540651673254 1.60509572884132
1.30374322638613 1.58331785634609
1.27606146151523 1.55781914079284
1.26235375858237 1.54395420468145
1.26228230665564 1.54490981027582
1.27510935366353 1.56075120044065
1.30003253726 1.59059140100894
1.33397325153109 1.62999260205069
1.36587409882271 1.66294621885062
1.3648586875966 1.64434062267312
1.34542019000646 1.60522004279896
1.30382821223971 1.58337679204158
1.27612310580354 1.55788403065767
1.26239006550295 1.54399135872947
1.26228939669277 1.54491344919932
1.27508588433952 1.56071999938891
1.29997870744922 1.59052567983134
1.33389525434277 1.62990121818167
1.36579710029422 1.66286555465244
1.36485554238365 1.64437741072037
};
\addlegendentry{Agent 2}
\addplot [green, dash pattern=on 1pt off 3pt on 3pt off 3pt, thick]
table {%
1.5 2
1.5 2
1.39206750477056 1.77330106329926
1.3139033504146 1.63348584924123
1.27588477550655 1.56862442579588
1.26618547020561 1.55112752052375
1.2758866760767 1.56159119939515
1.30004756305792 1.59035265586993
1.33391764916176 1.62977014887255
1.36586518108483 1.66285780021119
1.36482141367878 1.64423202754239
1.34536546248274 1.60510916101358
1.3037291241081 1.58332045597955
1.27605644931964 1.55782031548638
1.26235344208633 1.54395427722826
1.26228359305843 1.54491042084192
1.27511098123899 1.56075242541466
1.30003411398415 1.59059291718062
1.33397469549538 1.62999415769896
1.365875269818 1.66294748348379
1.36485902976331 1.64434072759864
1.34541820130289 1.60521676357859
1.30382596523038 1.58337375565377
1.27612072025615 1.55788379302082
1.26238987907946 1.54399130517675
1.26229040615731 1.54491418514208
1.27508730510639 1.56072133243888
1.29998018191174 1.5905272533811
1.33389665753283 1.62990279800772
1.36579825817412 1.66286682822169
1.36485588281773 1.64437752046642
};
\addlegendentry{Agent 3}
\addplot [red, dashed]
table {%
1.4 1.35
1.4 1.35
1.33136863731788 1.45814202672441
1.28768234739805 1.5061460064002
1.26634412070199 1.52570934313746
1.26339464420366 1.53963643648204
1.27534025539984 1.55973278746933
1.30007565542465 1.59068835828276
1.33401763345289 1.63025837085539
1.36593106879197 1.6631440107441
1.36485232200044 1.64435824715098
1.34537479941008 1.6051512904089
1.30373014102333 1.58333062988353
1.27605441130877 1.55782309611213
1.26235330311073 1.54395430517223
1.26228473286411 1.54491094774587
1.27511261748591 1.5607537777287
1.30003585464789 1.59059470317559
1.33397639460846 1.62999604936698
1.36587670649909 1.66294905913533
1.36485948138612 1.6443408948803
1.34541577152983 1.6052127297782
1.30382323805451 1.58337005269286
1.27611779467721 1.55788355031447
1.26238966122165 1.54399121250392
1.26229162122154 1.5449150471676
1.27508900491434 1.56072292935724
1.29998195566932 1.59052915923135
1.33389837038618 1.62990473727602
1.36579970004425 1.66286842027888
1.36485633790141 1.64437769499634
};
\addlegendentry{Agent 4}
\end{axis}

\end{tikzpicture}
    \caption{Evolution of the first two outputs of all agents. The eventual cooperative trajectory is marked with \textsf{x}.}
    \label{fig:synchronisation_positions}
\end{figure}
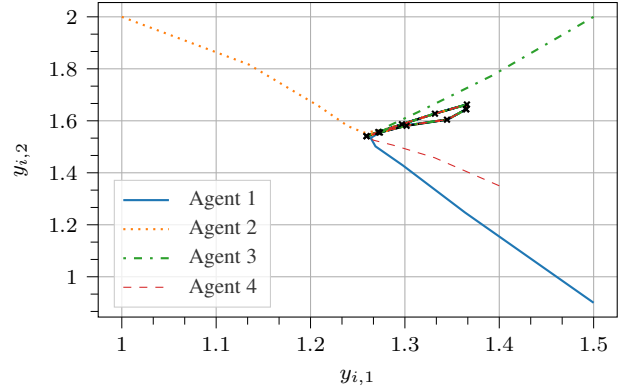%

%===============================================================================
\section{Conclusion}
We present a sequential distributed MPC scheme for multi-agent systems with periodic cooperative goals that can be characterised by a suitable set and associated cost.
Each agent uses a local tracking MPC with artificial output references that are penalised by the associated cost.
Constraint satisfaction and recursive feasibility are guaranteed for any agent independently of the actions of its neighbours.
Agents only need to communicate their optimal artificial output references once per time step; and a decentralised initialisation is possible.
We show that the cooperative goal is asymptotically achieved since agents move their artificial output references towards its accomplishment.
Future work will investigate the inclusion of coupling constraints, extending the framework to tasks that require, e.g. collision avoidance constraints or a maximum communication range.

\renewcommand*{\bibfont}{\scriptsize}
\bibliography{01_references}

\end{document}